%% file: moobench-cp.tex
\documentclass{gistt}

\usepackage[pdftex]{graphicx}
  \graphicspath{{./figs/}}
  \DeclareGraphicsExtensions{.pdf,.tex}
\usepackage{authblk}
\usepackage{todonotes}
\usepackage[hyperfirst=false,shortcuts]{glossaries}
\usepackage{siunitx}
\usepackage{paralist}
\usepackage{listings}
\usepackage[subrefformat=parens,labelformat=parens]{subfig}
\usepackage{xcolor}
\usepackage{tikz}
\usetikzlibrary{positioning,fit,calc}
\tikzset{block/.style={draw, text width=2.25cm ,minimum height=1.0cm, align=center},
line/.style={-latex}
}

\newcommand{\isrc}[1]{$\mathtt{#1}$}

\DeclareSIUnit{\million}{M}

\newglossaryentry{C++}{name={\makebox{C\hspace{-.04em}\raisebox{.15em}{\scriptsize\bf
+}\hspace{-.04em}\raisebox{.15em}{\scriptsize\bf +}}},description={C++}}
\newacronym{MooBench}{MooBench}{MooBench Observability Overhead Micro-Benchmark}

\lstdefinestyle{JavaStyle}{
    language=java,
    keywordstyle=\ttfamily\bfseries,
    numberstyle=\footnotesize,
    basicstyle=\ttfamily\fontsize{8}{8}\selectfont,
    numbers=left,
    numberstyle=\tiny\color{darkgray},
    showstringspaces=false,
    stepnumber=1,
    numbersep=5pt,
    tabsize=2,
    captionpos=b,
    breaklines=true,
    commentstyle=\tt\color{blue},
    xleftmargin=1.5em,
    framexleftmargin=1.5em,
    linewidth=\textwidth,
    fontadjust,
    emph={long,int,try,finally,return,if,else,do,while},
    emphstyle={\color{blue}},
}

\bibliography{moobench-cp.bib}

\title{Evaluating the Overhead of the Performance Profiler Cloudprofiler With MooBench}

\author[1]{Shinhyung Yang}
\author[2]{David Georg Reichelt}
\author[1]{Wilhelm Hasselbring}

\affil[1]{Kiel University, Kiel, Germany}
\affil[2]{Lancaster University Leipzig \& Universität Leipzig, Leipzig, Germany}
\affil[ ]{\{shinhyung.yang,hasselbring\}@email.uni-kiel.de,
          d.g.reichelt@lancaster.ac.uk}

\begin{document}

\maketitle

\begin{abstract}
Performance engineering has become crucial for the cloud-native
architecture. This architecture deploys multiple services, with each service representing an orchestration of containerized processes. 
OpenTelemetry is growing popular in the
cloud-native industry for observing the software's behaviour, and Kieker provides
the necessary tools to monitor and analyze the performance of target architectures.
Observability overhead is an important aspect of performance engineering and
MooBench is designed to compare different observability frameworks, including
OpenTelemetry and Kieker.

In this work, we measure the overhead of Cloudprofiler, a performance profiler
implemented in C++ to measure native and JVM processes. It minimizes the
profiling overhead by locating the profiler process outside the target process
and moving the disk writing overhead off the critical path with buffer blocks
and compression threads. Using MooBench, Cloudprofiler's buffered ID handler
with the Zstandard lossless data compression ZSTD showed an average execution
time of 2.28 microseconds. It is 6.15 times faster than the non-buffered and
non-compression handler.
\end{abstract}

\section{Introduction}

Cloud-native applications are best characterized by distributed services
provided to concurrent users, and deployment of microservices that scale
horizontally and vertically through container orchestration, often with
Kubernetes~\cite{Burns2016, Hasselbring2016, Gannon2017}.

Performance measurement of a cloud-native application is instrumental in
preventing failures and alleviating regressions such as prolonged response
times~\cite{APM2017}.  Ongoing efforts include the observability frameworks
Kieker~\cite{Hasselbring2020} and OpenTelemetry~\cite{Blanco2023}.
OpenTelemetry defines its data model with the Protocol Buffers serialization
framework\footnote{\url{https://protobuf.dev/}}, generating portable API and
data exports.  Kieker is an observability framework that instruments an
application for performance and behavioral analysis.

MooBench addresses another important dimension to the performance engineering
of cloud-native applications: performance overhead yielded by performance
observability frameworks~\cite{MooBench2015, Reichelt2021, Reichelt2024}.
It continuously evaluates the performance regression of its targets.
In empirical software engineering, benchmarks can be used for
comparing different methods, techniques and tools~\cite{EASE2021}. MooBench is designed for regression benchmarking within continuous integration pipelines~\cite{MooBench2015} of individual monitoring frameworks, not for comparing such frameworks against each other.

Cloudprofiler~\cite{Yang2023} can instrument both C/\ac{C++} and JVM-based
applications. Its use case includes profiling distributed matrix
computations~\cite{CMM2023}. Its logging interface is called \emph{handler},
and the identity~(ID) handler performs the disk I/O per each log entry.
The buffered ID handler reduces the overhead by writing log entries to memory
(buffer blocks) instead of disk.  The buffered and compressed ID handler
further exploits the performance by (1) redirecting buffers from the I/O thread
to parallel compression threads, which (2) increases the logs-per-second
I/O bandwidth.

In this paper, we use MooBench to evaluate the overhead of Cloudprofiler
instrumentation.  We added Cloudprofiler to MooBench.
MooBench evaluates five Cloudprofiler handlers: the null handler,
the non-buffered ID handler, the buffered and binary-encoded ID handler, and
two buffered and compressed ID handlers.
A buffered and compressed ID handler can be configured with one compression
codec, the Zstandard lossless data compression ZSTD codec, or the real-time
data compression LZO1X codec.
We present the results in Section~\ref{sec:eval}, and share the
code\footnote{\url{https://github.com/shinhyungyang/cloud_profiler}}
and datasets\footnote{\url{https://doi.org/10.5281/zenodo.13940072}}.

In the remainder of the paper, we
\begin{inparaenum}[(1)]
  \item describe how Cloudprofiler is incorporated into MooBench, and go over
  \item the experimental setup and
  \item results, 
  \item related work, and
  \item finalize in the conclusion section.
\end{inparaenum}

\section{Instrumenting with MooBench}

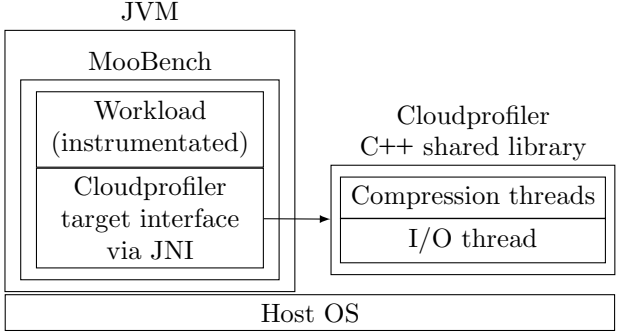
\begin{figure}[tbh]
\begin{tikzpicture}
\node[block, text width=2.75cm] (a) {Workload \\ (instrumentated)};
\node[block, text width=2.75cm, below=of a, yshift=10mm] (b) {Cloudprofiler \\
  target interface \\ via JNI};
\node[block, text width=3.25cm, minimum height=0cm, right=of b, yshift=-3.00mm] (d) {I/O thread};
\node[block, text width=3.25cm, minimum height=0cm, above=of d, yshift=-10.1mm] (e) {Compression threads};

\node[draw,inner xsep=2mm,inner ysep=1.5mm,fit=(a)(b),label={MooBench}](g){};
\node[draw, inner ysep=1.5mm, fit=(d)(e)] (c) {};
\node[fit=(d)(e), align=center, above=of c,yshift=-14.0mm] (k) {Cloudprofiler \\ \ac{C++}
  shared library};
\node[draw,inner xsep=2mm,inner ysep=4mm,yshift=2.5mm,fit=(a)(b)(g),label={JVM}](h){};
\node[inner xsep=0mm,inner ysep=0mm,yshift=0mm,fit=(c)(d)(h)](i){};
\node[block, text width=7.78cm, minimum height=0cm, below=of i, yshift=9.8mm](j){Host OS};

\draw[line] (b)-- (c);
\end{tikzpicture}
  \caption{Cloudprofiler Deployment in MooBench}
  \label{fig:cp-moobench}
\end{figure}

Cloudprofiler comprises two modules: the target
interface via JNI and the \ac{C++} shared library. Fig.~\ref{fig:cp-moobench}
depicts the Cloudprofiler modules within the MooBench architecture running as a
JVM process on the Host OS. The MooBench workload is instrumented at the source
code level with the Cloudprofiler interface. We incorporated the
instrumentation structure from~\cite{Reichelt2024}, which evaluated different
instrumentation technologies for JVM applications, including source code-level,
and bytecode-level instrumentations.

\begin{lstlisting}[label=lst:moobench-cp-instrumentation,
caption=\isrc{monitoredMethod} is instrumented to measure the execution time of
\isrc{extractedMethod}., style=JavaStyle, escapechar=|]
long monitoredMethod(long time, int depth) {
  cloud_profiler.logTS(ch_start, depth);
  try {
    return extractedMethod(time, depth);
  } finally {
    cloud_profiler.logTS(ch_end, depth);
  }
}

long extractedMethod(long time, int depth) {
  if (depth > 1) {
    return monitoredMethod(time, depth - 1);
  } else {
    long exitTime = System.nanoTime() + time;
    long curTime;
    do {
      curTime = System.nanoTime();
    } while (curTime < exitTime);
    return curTime;
  }
}
\end{lstlisting}

The \isrc{extractedMethod} function in
Listing~\ref{lst:moobench-cp-instrumentation} is MooBench's synthetic workload
that represents a trace: the \isrc{depth} of the trace and the \isrc{time}
spent are required.  The method is invoked recursively for the number of
\isrc{depth} and the last invocation returns after a duration of
\isrc{time}~\si{\nano\second}. MooBench targets without aspect weaving, e.g.,
Cloudprofiler, utilize \isrc{monitoredMethod} to manually instrument
\isrc{extractedMethod}. During execution, it deploys two Cloudprofiler
channels, \isrc{ch\_start} and \isrc{ch\_end}, which measures the span of
of \isrc{extractedMethod}.

\section{Experimental Setup}

We extended the  main branch of
MooBench\footnote{\url{https://github.com/kieker-monitoring/moobench}} to
incorporate the Cloudprofiler
framework.
We measured the overhead of three
frameworks: Cloudprofiler, Kieker for Java, and OpenTelemetry.

The benchmark is deployed on a bare metal server,
operated by Debian~12.6, running OpenJDK~17.0.2, and GCC~12.2.0. It
has two Intel Xeon~E5-2650 CPUs with eight physical cores on each, and
\SI{64}{\gibi\byte} RAM on each NUMA domain. It uses a
\SI{480}{\giga\byte} SSD.

We used MooBenchs default configurations, that includes \SI{2}{\million}
iterations,
where each iteration starts by calling \isrc{monitoredMethod} with \num{10} for
\isrc{depth}, and \num{0} for \isrc{time}. After an iteration, MooBench
collects the elapsed time, the garbage collection counts, and the current used
heap memory size. We report the results in Section~\ref{sec:eval}. The
measurement repeats 10 times and \SI{20}{\million} results are
collected in total.  Measuring an observability framework involves more than
one configuration, including non-instrumentation, deactivated instrumentation,
and other configurations specific to the target framework.  MooBench selects
\SI{10}{\million} execution results to create statistics for each configuration.

\paragraph{Cloudprofiler configurations}

We selected five configurations:
\begin{inparaenum}[(1)]
  \item the null handler,
  \item the non-buffered ID handler,
  \item the buffered and binary-encoded ID handler,
  \item the buffered and ZSTD-compressed ID handler, and
  \item the buffered and LZO1X-compressed ID handler.
\end{inparaenum}
The null handler is an empty JNI function without logging. A buffered handler
has 32 buffer blocks, where one block can
store up to \SI{1}{\million} log entries.
Four compression threads dequeue a non-compressed block
from a non-blocking multi-producer/multi-consumer~(MPMC) queue, and
enqueue a compressed block to another MPMC queue for the I/O thread.

\begin{figure*}[ht]
  \centering
    \scalebox{0.70}{
    \input{figs/evaluation.tex}
  }
  \scalebox{0.70}{
    \input{figs/evaluation-legend.tex}
  }
  \vspace*{-0.60em}
  \caption[MooBench Evaluation]
  {
    MooBench's evaluation comparison of Cloudprofiler, Kieker, and OpenTelemetry
  }
  \vspace*{-1.70em}
  \label{fig:evaluation}
\end{figure*}
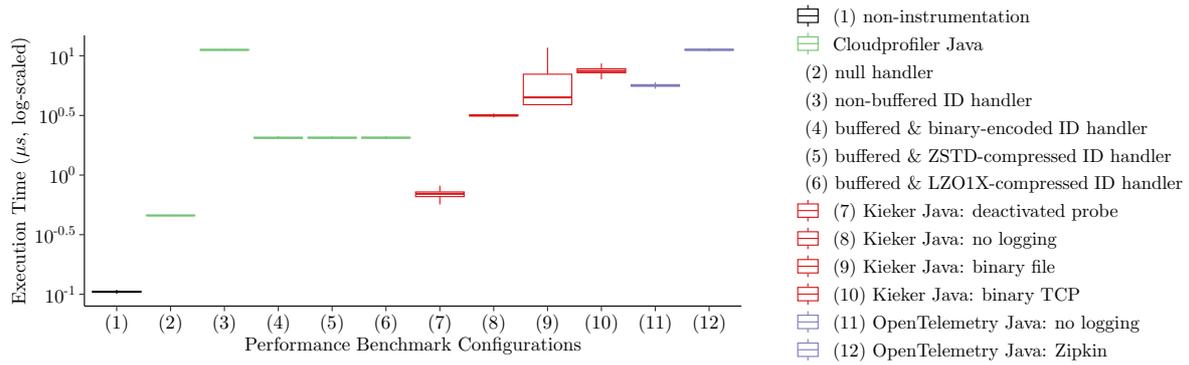

\section{Evaluation}\label{sec:eval}

We report the execution time results in Fig.~\ref{fig:evaluation}:
the non-instrumentation configuration~(1) is the baseline for other
configurations. Cloudprofiler's null handler~(2) performed
\SI{0.461}{\micro\second}, close to~(7) at \SI{0.717}{\micro\second}. The
buffered and binary-encoded ID handler performed \SI{2.294}{\micro\second}~(4),
and the buffered and compressed ID handlers performed
\SI{2.28}{\micro\second}~(5--6).  The Kieker configurations~(9, 10) exhibited
\SI{5.485}{\micro\second} and \SI{7.127}{\micro\second}, respectively.
Kieker and OpenTelemetry configurations rarely exhibited GC, less than
\num{10}~times during \SI{2}{\million} iterations. We used coefficient of variation
(CV) to observe the changes in the used heap memory size. Cloudprofiler,
Kieker, and OpenTelemetry exhibited maximum CVs of \SI{11.78}{\percent},
\SI{50.39}{\percent}, and \SI{54.35}{\percent}, respectively.

\section{Related Work}

Our work aims for measuring the observability overhead, which is the base for
overhead reduction. Eder et al.~\cite{eder2023comparison} compare the overhead
of distributed tracing in cloud environments. They use a microbenchmark that is
comparable to MooBench and find that Zipkins agent's overhead is lower
than OpenTelemetry's. Reichelt et al.~\cite{reichelt2023towards} aim for overhead
reduction by reducing the values created by Kieker. They find that the
overhead can be reduced from 4.77\,$\mu s$  to 0.39\,$\mu s$ (compared to a
baseline of 0.05\,$\mu s$).
Lengauer et al.~\cite{Lengauer2016} proposed a Java memory tracing techniques.
Each tracer thread may compress its buffer before enqueing, which will be
read by separate I/O threads.
Gebai and Dagenais~\cite{gebai2018tracers} compared the overhead of
system-level tracers with a tight loop microbenchmark that invokes a probe
function hooked to a tracepoint instrumentation in application code.

\section{Conclusion}

In this research, we incorporated Cloudprofiler into MooBench for regression
benchmarking, and evaluted its instrumentation overhead
differentiated by the use of memory buffers and parallel compression threads.
We intend to utilize MooBench for continuously benchmarking Cloudprofiler as
well as other frameworks.

\paragraph{Acknowledgment}
This research is funded by the Deutsche Forschungsgemeinschaft (DFG -- German
Research Foundation), grant no.~528713834.

\printbibliography

\end{document}

%% file: figs/evaluation.tex
\begin{tikzpicture}[x=1pt,y=1pt]
\definecolor{fillColor}{RGB}{255,255,255}
\path[use as bounding box,fill=fillColor,fill opacity=0.00] (0,0) rectangle (400.83,182.71);
\begin{scope}
\path[clip] (  0.00,  0.00) rectangle (400.83,182.71);
\definecolor{drawColor}{RGB}{255,255,255}
\definecolor{fillColor}{RGB}{255,255,255}

\path[draw=drawColor,line width= 0.6pt,line join=round,line cap=round,fill=fillColor] (  0.00,  0.00) rectangle (400.83,182.71);
\end{scope}
\begin{scope}
\path[clip] ( 44.09, 31.32) rectangle (395.33,177.21);
\definecolor{drawColor}{RGB}{0,0,0}

\path[draw=drawColor,line width= 0.6pt,line join=round] ( 61.36, 39.30) -- ( 61.36, 40.08);

\path[draw=drawColor,line width= 0.6pt,line join=round] ( 61.36, 38.77) -- ( 61.36, 37.95);
\definecolor{fillColor}{RGB}{255,255,255}

\path[draw=drawColor,line width= 0.6pt,fill=fillColor] ( 48.40, 39.30) --
	( 48.40, 38.77) --
	( 74.32, 38.77) --
	( 74.32, 39.30) --
	( 48.40, 39.30) --
	cycle;

\path[draw=drawColor,line width= 1.1pt] ( 48.40, 39.04) -- ( 74.32, 39.04);
\definecolor{drawColor}{RGB}{116,196,118}

\path[draw=drawColor,line width= 0.6pt,line join=round] ( 90.15, 80.26) -- ( 90.15, 80.53);

\path[draw=drawColor,line width= 0.6pt,line join=round] ( 90.15, 80.08) -- ( 90.15, 79.80);

\path[draw=drawColor,line width= 0.6pt,fill=fillColor] ( 77.20, 80.26) --
	( 77.20, 80.08) --
	(103.11, 80.08) --
	(103.11, 80.26) --
	( 77.20, 80.26) --
	cycle;

\path[draw=drawColor,line width= 1.1pt] ( 77.20, 80.14) -- (103.11, 80.14);

\path[draw=drawColor,line width= 0.6pt,line join=round] (118.94,169.56) -- (118.94,169.94);

\path[draw=drawColor,line width= 0.6pt,line join=round] (118.94,169.31) -- (118.94,168.92);

\path[draw=drawColor,line width= 0.6pt,fill=fillColor] (105.99,169.56) --
	(105.99,169.31) --
	(131.90,169.31) --
	(131.90,169.56) --
	(105.99,169.56) --
	cycle;

\path[draw=drawColor,line width= 1.1pt] (105.99,169.40) -- (131.90,169.40);

\path[draw=drawColor,line width= 0.6pt,line join=round] (147.73,122.27) -- (147.73,122.87);

\path[draw=drawColor,line width= 0.6pt,line join=round] (147.73,121.86) -- (147.73,121.24);

\path[draw=drawColor,line width= 0.6pt,fill=fillColor] (134.78,122.27) --
	(134.78,121.86) --
	(160.69,121.86) --
	(160.69,122.27) --
	(134.78,122.27) --
	cycle;

\path[draw=drawColor,line width= 1.1pt] (134.78,121.98) -- (160.69,121.98);

\path[draw=drawColor,line width= 0.6pt,line join=round] (176.52,122.27) -- (176.52,122.79);

\path[draw=drawColor,line width= 0.6pt,line join=round] (176.52,121.92) -- (176.52,121.38);

\path[draw=drawColor,line width= 0.6pt,fill=fillColor] (163.57,122.27) --
	(163.57,121.92) --
	(189.48,121.92) --
	(189.48,122.27) --
	(163.57,122.27) --
	cycle;

\path[draw=drawColor,line width= 1.1pt] (163.57,122.02) -- (189.48,122.02);

\path[draw=drawColor,line width= 0.6pt,line join=round] (205.31,122.28) -- (205.31,122.82);

\path[draw=drawColor,line width= 0.6pt,line join=round] (205.31,121.92) -- (205.31,121.36);

\path[draw=drawColor,line width= 0.6pt,fill=fillColor] (192.36,122.28) --
	(192.36,121.92) --
	(218.27,121.92) --
	(218.27,122.28) --
	(192.36,122.28) --
	cycle;

\path[draw=drawColor,line width= 1.1pt] (192.36,122.04) -- (218.27,122.04);
\definecolor{drawColor}{RGB}{215,25,28}

\path[draw=drawColor,line width= 0.6pt,line join=round] (234.10, 92.84) -- (234.10, 96.22);

\path[draw=drawColor,line width= 0.6pt,line join=round] (234.10, 90.33) -- (234.10, 86.09);

\path[draw=drawColor,line width= 0.6pt,fill=fillColor] (221.15, 92.84) --
	(221.15, 90.33) --
	(247.06, 90.33) --
	(247.06, 92.84) --
	(221.15, 92.84) --
	cycle;

\path[draw=drawColor,line width= 1.1pt] (221.15, 91.81) -- (247.06, 91.81);

\path[draw=drawColor,line width= 0.6pt,line join=round] (262.89,134.35) -- (262.89,135.14);

\path[draw=drawColor,line width= 0.6pt,line join=round] (262.89,133.81) -- (262.89,132.98);

\path[draw=drawColor,line width= 0.6pt,fill=fillColor] (249.94,134.35) --
	(249.94,133.81) --
	(275.85,133.81) --
	(275.85,134.35) --
	(249.94,134.35) --
	cycle;

\path[draw=drawColor,line width= 1.1pt] (249.94,134.06) -- (275.85,134.06);

\path[draw=drawColor,line width= 0.6pt,line join=round] (291.68,156.30) -- (291.68,170.58);

\path[draw=drawColor,line width= 0.6pt,fill=fillColor] (278.73,156.30) --
	(278.73,139.86) --
	(304.64,139.86) --
	(304.64,156.30) --
	(278.73,156.30) --
	cycle;

\path[draw=drawColor,line width= 1.1pt] (278.73,143.80) -- (304.64,143.80);

\path[draw=drawColor,line width= 0.6pt,line join=round] (320.47,159.18) -- (320.47,162.10);

\path[draw=drawColor,line width= 0.6pt,line join=round] (320.47,157.05) -- (320.47,153.51);

\path[draw=drawColor,line width= 0.6pt,fill=fillColor] (307.52,159.18) --
	(307.52,157.05) --
	(333.43,157.05) --
	(333.43,159.18) --
	(307.52,159.18) --
	cycle;

\path[draw=drawColor,line width= 1.1pt] (307.52,157.99) -- (333.43,157.99);
\definecolor{drawColor}{RGB}{128,125,186}

\path[draw=drawColor,line width= 0.6pt,line join=round] (349.26,150.67) -- (349.26,151.90);

\path[draw=drawColor,line width= 0.6pt,line join=round] (349.26,149.81) -- (349.26,148.48);

\path[draw=drawColor,line width= 0.6pt,fill=fillColor] (336.31,150.67) --
	(336.31,149.81) --
	(362.22,149.81) --
	(362.22,150.67) --
	(336.31,150.67) --
	cycle;

\path[draw=drawColor,line width= 1.1pt] (336.31,150.23) -- (362.22,150.23);

\path[draw=drawColor,line width= 0.6pt,line join=round] (378.05,169.70) -- (378.05,170.41);

\path[draw=drawColor,line width= 0.6pt,line join=round] (378.05,169.21) -- (378.05,168.47);

\path[draw=drawColor,line width= 0.6pt,fill=fillColor] (365.10,169.70) --
	(365.10,169.21) --
	(391.01,169.21) --
	(391.01,169.70) --
	(365.10,169.70) --
	cycle;

\path[draw=drawColor,line width= 1.1pt] (365.10,169.44) -- (391.01,169.44);
\end{scope}
\begin{scope}
\path[clip] (  0.00,  0.00) rectangle (400.83,182.71);
\definecolor{drawColor}{RGB}{0,0,0}

\path[draw=drawColor,line width= 0.6pt,line join=round] ( 44.09, 31.32) --
	( 44.09,177.21);
\end{scope}
\begin{scope}
\path[clip] (  0.00,  0.00) rectangle (400.83,182.71);
\definecolor{drawColor}{RGB}{0,0,0}

\node[text=drawColor,anchor=base west,inner sep=0pt, outer sep=0pt, scale=  1.03] at ( 22.90, 33.27) {10};

\node[text=drawColor,anchor=base west,inner sep=0pt, outer sep=0pt, scale=  0.72] at ( 33.15, 37.47) {-1};

\node[text=drawColor,anchor=base west,inner sep=0pt, outer sep=0pt, scale=  1.03] at ( 17.31, 65.40) {10};

\node[text=drawColor,anchor=base west,inner sep=0pt, outer sep=0pt, scale=  0.72] at ( 27.57, 69.59) {-0.5};

\node[text=drawColor,anchor=base west,inner sep=0pt, outer sep=0pt, scale=  1.03] at ( 25.29, 97.52) {10};

\node[text=drawColor,anchor=base west,inner sep=0pt, outer sep=0pt, scale=  0.72] at ( 35.55,101.72) {0};

\node[text=drawColor,anchor=base west,inner sep=0pt, outer sep=0pt, scale=  1.03] at ( 19.70,129.65) {10};

\node[text=drawColor,anchor=base west,inner sep=0pt, outer sep=0pt, scale=  0.72] at ( 29.96,133.85) {0.5};

\node[text=drawColor,anchor=base west,inner sep=0pt, outer sep=0pt, scale=  1.03] at ( 25.29,161.78) {10};

\node[text=drawColor,anchor=base west,inner sep=0pt, outer sep=0pt, scale=  0.72] at ( 35.55,165.97) {1};
\end{scope}
\begin{scope}
\path[clip] (  0.00,  0.00) rectangle (400.83,182.71);
\definecolor{drawColor}{gray}{0.20}

\path[draw=drawColor,line width= 0.6pt,line join=round] ( 41.34, 37.68) --
	( 44.09, 37.68);

\path[draw=drawColor,line width= 0.6pt,line join=round] ( 41.34, 69.80) --
	( 44.09, 69.80);

\path[draw=drawColor,line width= 0.6pt,line join=round] ( 41.34,101.93) --
	( 44.09,101.93);

\path[draw=drawColor,line width= 0.6pt,line join=round] ( 41.34,134.05) --
	( 44.09,134.05);

\path[draw=drawColor,line width= 0.6pt,line join=round] ( 41.34,166.18) --
	( 44.09,166.18);
\end{scope}
\begin{scope}
\path[clip] (  0.00,  0.00) rectangle (400.83,182.71);
\definecolor{drawColor}{RGB}{0,0,0}

\path[draw=drawColor,line width= 0.6pt,line join=round] ( 44.09, 31.32) --
	(395.33, 31.32);
\end{scope}
\begin{scope}
\path[clip] (  0.00,  0.00) rectangle (400.83,182.71);
\definecolor{drawColor}{gray}{0.20}

\path[draw=drawColor,line width= 0.6pt,line join=round] ( 61.36, 28.57) --
	( 61.36, 31.32);

\path[draw=drawColor,line width= 0.6pt,line join=round] ( 90.15, 28.57) --
	( 90.15, 31.32);

\path[draw=drawColor,line width= 0.6pt,line join=round] (118.94, 28.57) --
	(118.94, 31.32);

\path[draw=drawColor,line width= 0.6pt,line join=round] (147.73, 28.57) --
	(147.73, 31.32);

\path[draw=drawColor,line width= 0.6pt,line join=round] (176.52, 28.57) --
	(176.52, 31.32);

\path[draw=drawColor,line width= 0.6pt,line join=round] (205.31, 28.57) --
	(205.31, 31.32);

\path[draw=drawColor,line width= 0.6pt,line join=round] (234.10, 28.57) --
	(234.10, 31.32);

\path[draw=drawColor,line width= 0.6pt,line join=round] (262.89, 28.57) --
	(262.89, 31.32);

\path[draw=drawColor,line width= 0.6pt,line join=round] (291.68, 28.57) --
	(291.68, 31.32);

\path[draw=drawColor,line width= 0.6pt,line join=round] (320.47, 28.57) --
	(320.47, 31.32);

\path[draw=drawColor,line width= 0.6pt,line join=round] (349.26, 28.57) --
	(349.26, 31.32);

\path[draw=drawColor,line width= 0.6pt,line join=round] (378.05, 28.57) --
	(378.05, 31.32);
\end{scope}
\begin{scope}
\path[clip] (  0.00,  0.00) rectangle (400.83,182.71);
\definecolor{drawColor}{RGB}{0,0,0}

\node[text=drawColor,anchor=base,inner sep=0pt, outer sep=0pt, scale=  1.03] at ( 61.36, 19.31) {(1)};

\node[text=drawColor,anchor=base,inner sep=0pt, outer sep=0pt, scale=  1.03] at ( 90.15, 19.31) {(2)};

\node[text=drawColor,anchor=base,inner sep=0pt, outer sep=0pt, scale=  1.03] at (118.94, 19.31) {(3)};

\node[text=drawColor,anchor=base,inner sep=0pt, outer sep=0pt, scale=  1.03] at (147.73, 19.31) {(4)};

\node[text=drawColor,anchor=base,inner sep=0pt, outer sep=0pt, scale=  1.03] at (176.52, 19.31) {(5)};

\node[text=drawColor,anchor=base,inner sep=0pt, outer sep=0pt, scale=  1.03] at (205.31, 19.31) {(6)};

\node[text=drawColor,anchor=base,inner sep=0pt, outer sep=0pt, scale=  1.03] at (234.10, 19.31) {(7)};

\node[text=drawColor,anchor=base,inner sep=0pt, outer sep=0pt, scale=  1.03] at (262.89, 19.31) {(8)};

\node[text=drawColor,anchor=base,inner sep=0pt, outer sep=0pt, scale=  1.03] at (291.68, 19.31) {(9)};

\node[text=drawColor,anchor=base,inner sep=0pt, outer sep=0pt, scale=  1.03] at (320.47, 19.31) {(10)};

\node[text=drawColor,anchor=base,inner sep=0pt, outer sep=0pt, scale=  1.03] at (349.26, 19.31) {(11)};

\node[text=drawColor,anchor=base,inner sep=0pt, outer sep=0pt, scale=  1.03] at (378.05, 19.31) {(12)};
\end{scope}
\begin{scope}
\path[clip] (  0.00,  0.00) rectangle (400.83,182.71);
\definecolor{drawColor}{RGB}{0,0,0}

\node[text=drawColor,anchor=base,inner sep=0pt, outer sep=0pt, scale=  1.03] at (219.71,  7.49) {Performance Benchmark Configurations};
\end{scope}
\begin{scope}
\path[clip] (  0.00,  0.00) rectangle (400.83,182.71);
\definecolor{drawColor}{RGB}{0,0,0}

\node[text=drawColor,rotate= 90.00,anchor=base,inner sep=0pt, outer sep=0pt, scale=  1.03] at ( 12.57,104.27) {Execution Time ($\mu s$, log-scaled)};
\end{scope}
\end{tikzpicture}

%% file: figs/evaluation-legend.tex
\def\RectW{210.00}
\def\RectH{195.00}
\begin{tikzpicture}[x=1pt,y=1pt]
\definecolor{fillColor}{RGB}{255,255,255}
\path[use as bounding box,fill=fillColor,fill opacity=0.00] (0,0) rectangle (\RectW,\RectH);
\begin{scope}
\path[clip] (0,0) rectangle (\RectW,\RectH);
\definecolor{fillColor}{RGB}{255,255,255}

\path[fill=fillColor] (0,0) rectangle (\RectW,\RectH);
\end{scope}

\begin{scope}
\path[clip] (0,0) rectangle (\RectW,\RectH);

\definecolor{drawColor}{RGB}{255,255,255}
\definecolor{fillColor}{RGB}{255,255,255}
\path[draw=drawColor,line width= 0.6pt,fill=fillColor] 
                                        (0,0) rectangle (\RectW,\RectH);
\definecolor{drawColor}{RGB}{  0,  0,  0}
\definecolor{fillColor}{RGB}{255,255,255}

\path[draw=drawColor,line width= 0.6pt] (8.66,182.0) -- (8.66,184.16);
\path[draw=drawColor,line width= 0.6pt] (8.66,191.38) -- (8.66,193.56);
\path[draw=drawColor,line width= 0.6pt,fill=fillColor] 
                                        (3.26,184.16) rectangle (14.1,191.38);
\path[draw=drawColor,line width= 0.6pt] (3.26,187.78) -- (14.1,187.78);

\definecolor{drawColor}{RGB}{116,196,118}
\definecolor{fillColor}{RGB}{255,255,255}

\path[draw=drawColor,line width= 0.6pt] (8.66,167.0) -- (8.66,169.16);
\path[draw=drawColor,line width= 0.6pt] (8.66,176.38) -- (8.66,178.56);
\path[draw=drawColor,line width= 0.6pt,fill=fillColor] 
                                        (3.26,169.16) rectangle (14.1,176.38);
\path[draw=drawColor,line width= 0.6pt] (3.26,172.78) -- (14.1,172.78);

\definecolor{drawColor}{RGB}{215,25,28}
\definecolor{fillColor}{RGB}{255,255,255}

\path[draw=drawColor,line width= 0.6pt] (8.66,77.0) -- (8.66,79.16);
\path[draw=drawColor,line width= 0.6pt] (8.66,86.38) -- (8.66,88.56);
\path[draw=drawColor,line width= 0.6pt,fill=fillColor] 
                                        (3.26,79.16) rectangle (14.1,86.38);
\path[draw=drawColor,line width= 0.6pt] (3.26,82.78) -- (14.1,82.78);

\path[draw=drawColor,line width= 0.6pt] (8.66,62.0) -- (8.66,64.16);
\path[draw=drawColor,line width= 0.6pt] (8.66,71.38) -- (8.66,73.56);
\path[draw=drawColor,line width= 0.6pt,fill=fillColor] 
                                        (3.26,64.16) rectangle (14.1,71.38);
\path[draw=drawColor,line width= 0.6pt] (3.26,67.78) -- (14.1,67.78);

\path[draw=drawColor,line width= 0.6pt] (8.66,47.0) -- (8.66,49.16);
\path[draw=drawColor,line width= 0.6pt] (8.66,56.38) -- (8.66,58.56);
\path[draw=drawColor,line width= 0.6pt,fill=fillColor] 
                                        (3.26,49.16) rectangle (14.1,56.38);
\path[draw=drawColor,line width= 0.6pt] (3.26,52.78) -- (14.1,52.78);

\path[draw=drawColor,line width= 0.6pt] (8.66,32.0) -- (8.66,34.16);
\path[draw=drawColor,line width= 0.6pt] (8.66,41.38) -- (8.66,43.56);
\path[draw=drawColor,line width= 0.6pt,fill=fillColor] 
                                        (3.26,34.16) rectangle (14.1,41.38);
\path[draw=drawColor,line width= 0.6pt] (3.26,37.78) -- (14.1,37.78);

\definecolor{drawColor}{RGB}{128,125,186}
\definecolor{fillColor}{RGB}{255,255,255}

\path[draw=drawColor,line width= 0.6pt] (8.66,17.0) -- (8.66,19.16);
\path[draw=drawColor,line width= 0.6pt] (8.66,26.38) -- (8.66,28.56);
\path[draw=drawColor,line width= 0.6pt,fill=fillColor] 
                                        (3.26,19.16) rectangle (14.1,26.38);
\path[draw=drawColor,line width= 0.6pt] (3.26,22.78) -- (14.1,22.78);

\path[draw=drawColor,line width= 0.6pt] (8.66,2.0) -- (8.66,4.16);
\path[draw=drawColor,line width= 0.6pt] (8.66,11.38) -- (8.66,13.56);
\path[draw=drawColor,line width= 0.6pt,fill=fillColor] 
                                        (3.26,4.16) rectangle (14.1,11.38);
\path[draw=drawColor,line width= 0.6pt] (3.26,7.78) -- (14.1,7.78);

\definecolor{drawColor}{RGB}{0,0,0}

\node[text=drawColor,anchor=base west,inner sep=0pt, outer sep=0pt, scale= 1.0]
  at (21.99, 184.21)  {(1) non-instrumentation};
\node[text=drawColor,anchor=base west,inner sep=0pt, outer sep=0pt, scale= 1.0]
  at (21.99, 169.21)  {Cloudprofiler Java};
\node[text=drawColor,anchor=base west,inner sep=0pt, outer sep=0pt, scale= 1.0]
  at (7.0, 154.21)  {(2) null handler};
\node[text=drawColor,anchor=base west,inner sep=0pt, outer sep=0pt, scale= 1.0]
  at (7.0,  139.21) {(3) non-buffered ID handler};
\node[text=drawColor,anchor=base west,inner sep=0pt, outer sep=0pt, scale= 1.0]
  at (7.0,  124.21) {(4) buffered \& binary-encoded ID handler};
\node[text=drawColor,anchor=base west,inner sep=0pt, outer sep=0pt, scale= 1.0]
  at (7.0,  109.21) {(5) buffered \& ZSTD-compressed ID handler};
\node[text=drawColor,anchor=base west,inner sep=0pt, outer sep=0pt, scale= 1.0]
  at (7.0,  94.21) {(6) buffered \& LZO1X-compressed ID handler};
\node[text=drawColor,anchor=base west,inner sep=0pt, outer sep=0pt, scale= 1.0]
  at (21.99, 79.21) {(7) Kieker Java: deactivated probe};
\node[text=drawColor,anchor=base west,inner sep=0pt, outer sep=0pt, scale= 1.0]
  at (21.99, 64.21)  {(8) Kieker Java: no logging};
\node[text=drawColor,anchor=base west,inner sep=0pt, outer sep=0pt, scale= 1.0]
  at (21.99, 49.21)  {(9) Kieker Java: binary file};
\node[text=drawColor,anchor=base west,inner sep=0pt, outer sep=0pt, scale= 1.0]
  at (21.99, 34.21)  {(10) Kieker Java: binary TCP};
\node[text=drawColor,anchor=base west,inner sep=0pt, outer sep=0pt, scale= 1.0]
  at (21.99, 19.21)  {(11) OpenTelemetry Java: no logging};
\node[text=drawColor,anchor=base west,inner sep=0pt, outer sep=0pt, scale= 1.0]
  at (21.99, 4.21)  {(12) OpenTelemetry Java: Zipkin};
\end{scope}
\end{tikzpicture}